\documentclass[prl,aps,twocolumn,showpacs]{revtex4}
\usepackage{graphicx}
\usepackage{amsmath}

\newcommand \beq  {\begin{equation}}
\newcommand \eeq  {\end{equation}}
\newcommand \bea {\begin{eqnarray} }
\newcommand \eea {\end{eqnarray}}

\begin{document}
\title{Quantum dot in the pseudogap Kondo state}
\author{J. Hopkinson, K. Le Hur, and {\'E}. Dupont}
\affiliation{D{\'e}partement de Physique, Universit{\'e} de Sherbrooke, Sherbrooke, Qu{\'e}bec, Canada}

\begin{abstract}
We investigate the transport properties of a (small) quantum dot connected 
to Fermi liquid leads with a power-law density of states (DOS).  
Such a system, if experimentally realizable, will have interesting physical properties including: (i) non-saturating Coulomb blockade peak widths; (ii) a non-unitary Kondo peak symmetrically placed between Coulomb blockade peaks; (iii) an absence of conductance away from particle-hole symmetry at sufficiently low temperatures; and (iv) evidence of a quantum critical point as a function of dot-lead hopping. These properties are compared and contrasted with one dimensional Luttinger 
systems exhibiting a power-law ``tunneling-DOS''.
\end{abstract} 


\maketitle

The realization{\cite{raikh}} and verification{\cite{goldhaber}} that single-impurity Kondo physics is experimentally accessible in quantum dot systems led to a renaissance of interest in this problem, inspiring works as diverse as investigations of non-equilibrium effects{\cite{rosch}} to the observation of many-body resonances such as the quantum corral{\cite{manoharan}}.  The role of electron fractionalization has been important to descriptions of 
the fractional quantum Hall effect and spin-charge separation in one dimension (1D).
  We study the transport properties of a (small) quantum dot connected
to Fermi Liquid (FL) leads with a normalized power-law density of states (DOS) $\rho(\epsilon) = \frac{r+1}{2D}|\frac{\epsilon}{D}|^r$, $D$ being the bandwidth, which allows us 
to probe exotic physics of the
 pseudogap Kondo model{\cite{withoff}}  at the nanoscale.
This may allow for the first measurement of a Kondo state with a fractional phase shift of its conduction electrons{\cite{chen,gonzalez}}.  In contrast with the constant DOS case, odd Coulomb valleys no longer fill in, so Coulomb blockade peaks (Cbps) are well separated from Kondo peaks due to an opacity to conductance introduced by particle-hole symmetry (p-hs) breaking terms which completely change the nature of the strong coupling fixed point{\cite{gonzalez}}.
Preliminary results can be found in Ref. \onlinecite{John}.  
Can such a power-law DOS be realized? 
One is naturally drawn towards materials 
with nodal quasiparticles (qps) along their Fermi surfaces such as the d-wave high temperature cuprate superconductors (d-sc) or heavy fermion systems. The Kondo effect we describe will not be measureable when $r=1$ excluding immediately 
the former which exhibit a linear DOS along the nodal directions. 
However, it is reasonable to expect realizations of such ideas in the future.

When tunneling through a single-barrier, one of the main effects of interactions in 1D Luttinger systems is to
renormalize the {\it tunneling}-DOS (TDOS), which means the DOS to add an electron at an energy $\epsilon$ {\cite{egger}:
$\bar{\rho}(\epsilon)\propto{|\epsilon|}^{-1+1/g}$, $g$ is the well-known Luttinger exponent and $g<1$ for repulsive interactions.
A similar effect can be 
obtained in the case of a mesoscopic conductor embedded in an electrical circuit 
with an ohmic resistance $R$. Indeed, by tunneling through a tunnel junction in the presence of an ohmic environment, in the linear response r{\'e}gime the theory predicts a conductance $G(V)\propto 
|V|^{2R/R_K}$, $R_K$=$\frac{h}{e^2}$=25.8k$\Omega$ being the quantum of resistance
and $V$ the bias voltage {\cite{ingold}}, which from Fermi's golden rule $G(V)\propto \bar{\rho}(V)^2$ might also be interpreted as a power-law TDOS $\bar{\rho}(\epsilon)\propto {|\epsilon|}^{R/R_K}$. The 
mapping between these two problems has been addressed in Ref. {\onlinecite{kanef}} and explicitly
proven recently in Ref. {\onlinecite{safi}}.  Evidence for such a small power-law TDOS has been shown in different materials including small-capacitance junctions{\cite{cleland}}, (multiwall) nanotubes{\cite{bach}}, and NbSe$_3$ quantum wires{\cite{slot}}; the last two might find description in terms of a multi-mode Luttinger theory without single-particle hopping between modes{\cite{matveev}}. By identifying $\bar{\rho}(\epsilon) = {\rho}(\epsilon)$, 
one may wonder to what extent an analogy between the conductance 
of a FL with a power-law DOS and that of a 1D system with a power-law TDOS holds.

Here we focus on a quantum dot coupled to FL leads possessing a power-law DOS and compare and contrast with the situation of a quantum dot coupled to 1D leads.

Kane and Fisher{\cite{kanef}} realized that, while a single impurity in a Luttinger liquid is localizing, a second barrier restores the ability of the system to conduct as $T\rightarrow$ 0, with $G=gG_0$, where the Luttinger parameter $g<1$ for repulsive interactions and $G_0$ is the unitary conductance $\frac{2e^2}{h}$.  Furusaki and Nagaosa{\cite{furusaki}} extended this 1D work to extract the temperature dependence of the height of the Cbps, found to grow as $T^{\frac{1}{g}-2}$ at low temperatures, and the width of the peak, found to vanish as $T^{\frac{1}{g}-1}$ ($T$, for long range interactions), with experimental support{\cite{ausl}}.  
Recently, Nazarov and Glazman{\cite{nazarov}} revisited the resonant tunneling problem in 1D to build a non-perturbative theory of the conductance valid
in a broad region of $T$. The strong interaction limit ($g=1/2$) has been treated to similar effect{\cite{komnik}}. We find the width of the Cbps is at low $T$ governed by the power of the DOS, vanishing as $T^{r}$.  The height of the Cbps and Kondo peaks asymptote to $G$=$G_0\cos^2(\frac{\pi r}{2})$ as $T\rightarrow 0$.   Kondo physics is predicted to occur when $r<\frac{1}{2}$, the coupling $J$ is sufficiently large, and the number of dot electrons is odd, results different from 1D as detailed in Fig. {\ref{figure3}}.  The potential for a tunable quantum critical point (qcp) and implications at asymmetric points are presented. 

We consider the small dot, weak tunneling limit in which the dot is a collection of discrete levels of average spacing $\delta E$, of similar size as the charging energy, $E_c$ (${\mathcal{O}}$($\frac{e^2}{2C}$)), of the dot ($e$: electron charge; $C$: total capacitance of quantum dot).  The energy to add/subtract an electron from the dot is $E^{\pm}_c$ = $E_{tot}$($V_G$,$n+1$) - $E_{tot}$($V_G$, $n$)  = ($n+\frac{1}{2}$ - $\frac{C_G V_G}{e}$)$\frac{e^2}{C}$, where $n$ is the initial(final) number of electrons on the dot for +(-), $E_{tot}$ is the total dot energy, and $V_G$ is an external gate voltage coupled capacitively ($C_G$) to the dot.  For temperatures satisfying $\delta E<T<E_c$, the physics of the dot is dominated by Coulomb blockade, whereas for $T<E^*$, single-level Coulomb blockade ($E^*\equiv\min(\delta E, E_c)$) occurs.
Thus, for $T<E^*$, we can describe an effective hopping across the dot in terms of a single impurity
Anderson model{\cite{raikh}},
\begin{eqnarray}
H&=&\sum_{k\alpha\sigma}\left(\epsilon_k c_{k\alpha\sigma}^{\dagger} c_{k\alpha\sigma} + (t_{k\alpha}c_{k\alpha\sigma}^{\dagger}d_{\sigma} + h.c.)\right) + \epsilon_dn_d \nonumber \\& &+ Un_{d\downarrow}n_{d\uparrow}, \label{equation1}
\end{eqnarray}  
where $c_{k\alpha\sigma}$ destroys a conduction electron of momentum $k$, spin $\sigma=\uparrow,\downarrow$, from lead $\alpha=L,R$ ($L$: left and $R$: right). Moreover, $\epsilon_d$
denotes the energy of the highest occupied level on the dot with occupancy $n_d$ = $d_{\sigma}^{\dagger}d_{\sigma}$ and $U$=($E^{+}_c$+$E^{-}_c$)
takes into account the Coulomb repulsion on the dot.
Keep in mind that here the conduction band is embodied by 
a power law DOS $\rho(\epsilon) = \frac{r+1}{2D}|\frac{\epsilon}{D}|^r$.

\begin{figure}
\includegraphics[scale=0.25]{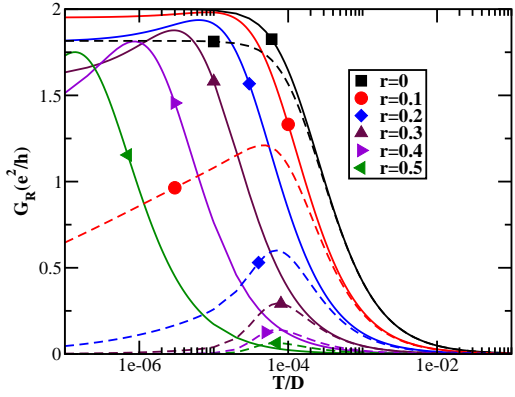}
\caption{\label{figure1}Resonant tunneling conductance $G_R$ vs. temperature $T$ for tunneling, $t=0.01D$.  At $\epsilon_d = 0$ (solid) all curves rise toward the unitary limit before saturating at $\frac{2e^2}{h}\cos^2(\frac{\pi r}{2})$.  At small $\epsilon_d=10^{-4}D$ (dashed), curves vanish as $T^{2r}$ as $T\rightarrow 0$.}
\end{figure}

{\it{Resonant level limit:}} The resonant limit is reached by tuning the gate voltage $V_G$ such that the energy to add/subtract one electron obeys $\delta E^{\pm}_c=0${\cite{raikh}}.  In this case the effective Coulomb interaction vanishes ($U=0$) and the physics is that of a single level of energy $\epsilon_d$.  To derive an expression for the conductance of the quantum dot system it is helpful to consider the current incident on, $I_L$, and transmitted from, $I_R$, the dot.  If we first assume the dot is transparent then it is simple to write $I_L = n_L e v = \frac{2e}{l}\sum_k f_{L}(\epsilon_k) v = \frac{2e}{h}\int_{-\infty}^{\infty}d\epsilon f_{L}(\epsilon)$ and $I = I_L - I_R$.  Here we assume an infinitesimal voltage drop from left to right across the dot and define $n_{L(R)} = \frac{N_{L(R)}}{l}$  and $f_{L(R)}(\epsilon)$ as the density of particles and the Fermi function of the left(right) lead respectively; $l$ the length of the lead; and $v=\frac{1}{\hbar} \frac{\partial \epsilon}{\partial k}$ the drift velocity of the particles.  In the first step we replaced $\int \rho(\epsilon) d\epsilon \rightarrow \sum_k$ followed in the second by the replacement $\sum_k \rightarrow \frac{l}{2\pi}\int dk$.  This derivation is appropriate for FL leads (with electron-like
quasiparticles).
For non-unitary transmittance an additional factor occurs within the integral due to elastic scattering from the double barrier{\cite{wingreen}} to yield, 
\begin{equation}
I = \frac{2e}{h}\int_{-D}^{D} d\epsilon
\frac{\left(f_L(\epsilon)-f_R(\epsilon)\right)4\Gamma_L(\epsilon)\Gamma_R(\epsilon)}{(\epsilon-\epsilon_d+\Lambda(\epsilon))^2 + \left(\Gamma_L(\epsilon) + \Gamma_R(\epsilon)\right)^2}.
\end{equation} 
\begin{figure}
\includegraphics[scale=0.32]{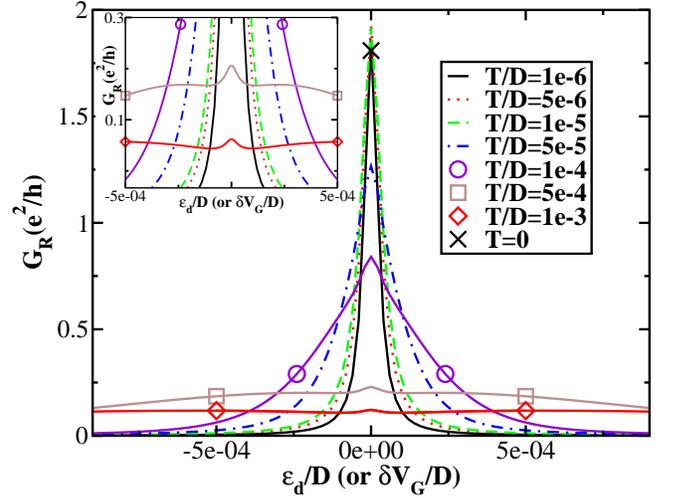}
\caption{\label{figure2}Temperature evolution of the conductance near the resonant point ($\epsilon_d=0$, $\delta V_G=0$) for $r=0.2$, $t=0.01D$.  At high temperatures unlike for $r=0$, there is a clear minimum (seen magnified in the inset) as a function of $V_G$ and the maximum of the conductance does not occur at resonance.  Further lowering the temperature, the conductance at $\epsilon_d=0$ grows as in Fig. 1, while the weight in the tails decreases.}
\end{figure}
Here, we approximate $t_{k\alpha}=t_{\alpha}$, $\Gamma_L(\epsilon)$ and $\Gamma_R(\epsilon)$ are the widths
of the quasilocal level associated with escape to respectively the left-hand and right-hand leads and $\Lambda(\epsilon)=\Lambda_L(\epsilon)+\Lambda_R(\epsilon)$ with $\Lambda_{\alpha}(\epsilon)$ the corresponding real parts of the self-energies.
These are defined as $\Gamma_{\alpha}(\epsilon)=\pi t_{\alpha}^2\rho(\epsilon)${\cite{raikh}} and $\Lambda_{\alpha}(\epsilon)=-t_{\alpha}^2\rho(\epsilon)sgn(\epsilon)\times$ $((\frac{|\epsilon|}{D})^{1-r}\frac{1}{r-1}{}_2F_1(1,\frac{1-r}{2};\frac{3-r}{2};(\frac{\epsilon}{D})^2)$$+\pi\tan(\frac{\pi r}{2}))$.\cite{note} 
Below, we consider symmetric barriers where $t_{\alpha}=t$ which leads to 
$\Gamma_{L(R)}(\epsilon) = \Gamma_0|\frac{\epsilon}{D}|^r$ with $\Gamma_0$=$\pi\frac{(r+1)}{2D}t^2$.  The difference between $E_c$ and $D$ may be quite large, as $E_c \approx 1 K$, whereas $D$ is the bandwidth of the electron leads.  Using (2) we can calculate the bare conductance ($G_R$ = $\frac{dI}{dV}|_{V\rightarrow 0}$ assuming $\mu_{L(R)}=\pm \frac{eV}{2}$) to arrive at, 
\begin{equation}
\hskip-0.5pc G_R=\frac{2e^2}{h} \frac{\gamma ^2 (r)}{k_BT} \int_{-D}^{D}  d\epsilon \frac{|\epsilon|^{2r} f(\epsilon) (1-f(\epsilon))}{(\epsilon-\epsilon_d+\Lambda(\epsilon))^2 + \gamma ^2 (r) |\epsilon|^{2r}}, 
\end{equation}
where $\gamma ^2 (r) =   \frac{\pi^2 t^4 (r+1)^2}{D^{2(1+r)}}$, $f(\epsilon)$: Fermi function at $V=0$.
It is straightforward to take the $T=0$ limit of this expression at resonance ($\epsilon_d=0$) to find: $G_R$=$\frac{2e^2\cos^2(\frac{\pi r}{2})}{h}$ for $r<1$; $G_R=0$ if $r\ge1$.  The conductance through a quantum dot can also be found{\cite{raikh}} as $G=G_0\sin^2(\delta)$ where $\delta(\epsilon)= \frac{\pi}{2}(1-r sgn(-\epsilon))$ is known (for $r<1$) to be the phase shift of the $U=0$ Anderson model{\cite{gonzalez,chen}}. The fractional phase shift can be interpreted as a decoupling of some fraction, $r$, of spins of the conduction electrons at $\epsilon_F$.  The temperature below which this saturation occurs is a decreasing function of $r$ as shown in Fig. \ref{figure1}.  
Away from resonance ($\epsilon_d \ne 0$), at high temperatures the curves follow those of the resonant case, and for $r=0$ saturate to a non-zero value determined by the distance away from the resonant point with a Lorentzian lineshape of width $\Gamma_0$.  For $r\ne 0$, the high temperature curves again follow those at resonance but exhibit a maximum which gradually crosses over at very low temperatures to the power-law form $T^{2r}$, such that one is left at $T=0$ with $\delta$ function peaks as a function of $\epsilon_d$ in place of the Lorentzian seen when $r=0$.  

We plot the resonant lineshapes as a function of $\delta V_G$ as one decreases the temperature in Fig. \ref{figure2} for the case $r=0.2$.  One observes a distinctly non-Lorentzian shape to these curves at high temperatures where a double-peak structure is evident--the wider peak exhibiting a pseudogap-like behavior as $V_G \rightarrow 0$ to effectively shift the conductance maximum away from 0 as seen in the inset.  As the temperature decreases, the central peak grows to eventually dwarf this outer structure.  We stress that even this central peak does not have a Lorentzian line-width as can be seen by plotting the half-width vs. temperature (not shown), which is seen for $r=0$ to saturate to the value $\Gamma_0$ and to vanish approximately as $T^{r}$ at sufficiently low temperatures (all curves have approximately linear $T$-dependence at high temperatures).  Here, we still note some similarities with 1D systems with a TDOS{\cite{kanef,furusaki}; this suggests that in 
the case of two symmetric barriers in 1D, for certain ranges of $\epsilon$ and $T$, the model may be also rewritten in terms of decoupled elastic scattering amplitudes $\Gamma_{L(R)}(\epsilon) = \pi t^2\bar{\rho}(\epsilon)$ with $\bar{\rho}(\epsilon)\propto {|\epsilon|}^{-1+1/g}$ being the TDOS at each barrier, as emphasized in Ref.{\cite{nazarov}}.

{\it{Kondo limit:}}
We have seen that a power-law DOS enhances Coulomb blockade to such an extent that, at zero temperature, Lorentzian lineshapes have been replaced by delta-function peaks about the resonant points where the energy to add or subtract an electron $E_c^{+(-)}$ vanishes.  In regular quantum dots ($r=0$), Coulomb valleys possessing an odd number of electrons allow certain spin-flip processes which at sufficiently low energies $T<E^*$, grow to eventually allow unitary conductance.  Can this same physics be realized when the DOS of the conduction electrons in the leads follows a power law?  To address this question we consider, for the same r{\'e}gime of $T$, the close vicinity 
of the point midway between two resonant peaks, with an odd number of electrons in the dot.  This allows us to restrict the energy levels to a filled level, with occupancy $n_d$ = $d_{\sigma}^{\dagger}d_{\sigma}$, $E_c^-$ = $\frac{e^2}{2C}\equiv -\epsilon_d$ below the Fermi energy and an unfilled level $E_c^+$ = $\frac{e^2}{2C}\equiv U + \epsilon_d$ above. 

\begin{figure}
\includegraphics[scale=0.4]{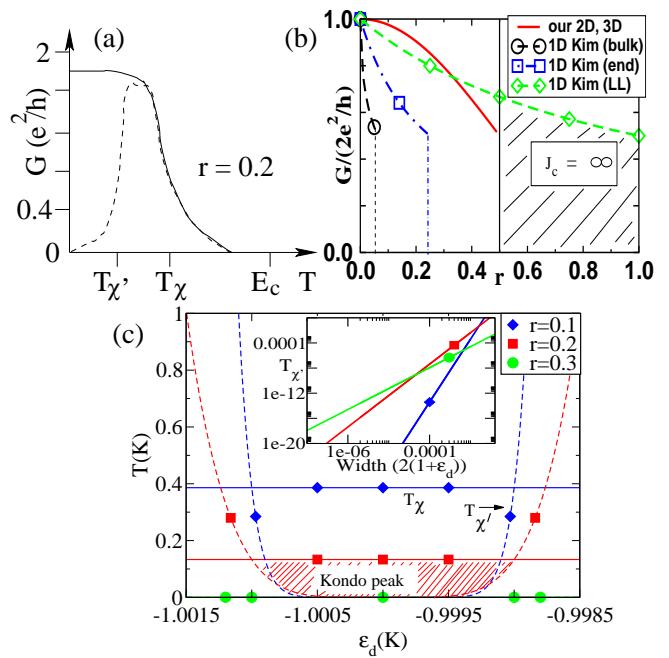}
\caption{\label{figure3} (a) Sketch of the conductance for $r=0.2$ in a Kondo valley: (solid) at p-hs $U=-2\epsilon_d$ (onsets below $T_{\chi}$); (dashed)  for $V\ne 0$ vanishes below $T_{\chi'}$. (b) The maximal height of the Kondo peak (at $U=-2\epsilon_d$) showing the $\cos^2(\frac{\pi}{2}r)$ dependence and comparison with the 1D case at the one-channel Kondo fixed point $G=\frac{2e^2g}{h}${\cite{kim}}  for:  Luttinger liquids $g=1/(r+1)$; carbon nanotubes tunneling through the end{\cite{egger}} $g_{end} = 1/(4r + 1)$; the bulk{\cite{egger}} $g_{bulk}=1+4r -\sqrt{8r+16r^2}$.  (c) The $V_G$ dependence of the crossover energy scales to enter the Kondo r{\'e}gime if $t=31.6K$, $\delta E \approx$ $E_c=1K$ and $D=10^3K$; $T_{\chi}$: (solid); $T_{\chi'}$: (dashed).  The filled region labelled Kondo peak delineates the boundaries of the $r=0.2$ Kondo effect or the width of the conductance peak
in the Kondo r\'egime located in between the resonant peaks at $\epsilon_d$=$0$ and $\epsilon_d$=$-2K$. Note that, for the parameter choices indicated, $T_{\chi}\rightarrow0$ for $r\gtrsim 0.31$. (inset) a log log plot: $T_{\chi'}$ vs width of the Kondo region at low $T$ for different values of $r$ (a $\delta$-function at $T=0$).}
\end{figure}
In the limit $-\epsilon_d$ ($n$ odd dot energy), $U$ + $\epsilon_d$ (energy of 1st excitation) $>>$ ($\Gamma_0$, $T$) we resort to the Schrieffer-Wolff transformation to
obtain the Kondo model{\cite{raikh}}
\begin{equation}
H_K = \sum_{k,k',\mu,\nu}\left(J_{k,k'}c_{k\mu}^{\dagger}\frac{\sigma_{\mu\nu}}{2}c_{k'\nu}S + V_{k,k'}c_{k\mu}^{\dagger}c_{k'\mu}\right),
\end{equation} 
where $S$ is the impurity spin of the level,
 the Kondo coupling $J_{k,k'}$ = 2($\frac{1}{|\epsilon_d|} + \frac{1}{|U+\epsilon_d|})t_k t_{k'}$, and the potential scattering $V_{k,k'}$ = $\frac{1}{2} (\frac{1}{|\epsilon_d|}- \frac{1}{|U+\epsilon_d|})t_k t_{k'}$ vanishes at the symmetric point $U=-2\epsilon_d$.  The lead index has disappeared as only the symmetric combination of leads is coupled to the level through Eq.(\ref{equation1}) {\cite{raikh}}.
For FL leads, it has been shown{\cite{raikh}} that within the Kondo r{\'e}gime scattering through the impurity spin of the dot simply introduces a phase shift of the conduction electrons.  Thus it is possible to map the strong coupling r{\'e}gime of the Anderson model to the $U=0$ resonant Anderson model.  For a constant density of states in the lead ($r=0$), the strong coupling fixed point of the Kondo model corresponds to a phase shift $\delta = \pi /2$ so one recovers unitary conductance as $T \rightarrow 0$.   Below, we treat two cases arising when $r\ne 0$ using knowledge of the pseudogap Kondo model{\cite{withoff,chen,gonzalez}}.  

For p-hs ($V$$=$$V_{k,k'}$=0) when the dot-lead hybridization is sufficiently large one can once more perform this mapping to the pseudogap $U=0$ resonant Anderson model{\cite{gonzalez}}. Hence, provided $J = J_{k,k'}$ is greater than a critical value $J_c$, as $r$ decreases the conductance at the point symmetrically placed between Cbps should exhibit a low temperature rise in conductivity to reach $G=G_R$ at $T=0$.   
The value of $J_c$ can be estimated following the poor man's scaling analysis of Ref. {\cite{withoff}}
and in our case we find $J_c\approx \frac{2rD}{r+1}|\frac{D}{E^*}|^r$ ($\Gamma_{0c} \sim \frac{\pi r E_c}{4}|\frac{D}{E^*}|^r$) for small r.  Numerical work in the $D=E^*$ limit{\cite{ingersent,gonzalez}} shows $J_c$ ($\Gamma_{oc}$) diverges at $r=0.5$.  Nodal qps of d-sc leads ($r=1$) should not support conductance. As a function of the matrix elements governing the hopping between dot and leads a qcp exists. For $J<J_c$, the local moment is unscreened blocking transport through the quantum dot $(G = 0)$.    For $J>J_c$ one enters{\cite{withoff}} the Kondo partially screened r{\'e}gime below the Kondo scale $T_{\chi} \sim E^*(\frac{J-J_c}{J})^{\frac{1}{r}}${\cite{withoff}} leading to $G = G_0\cos^2(\frac{\pi}{2}r)$.
Close to p-hs ($V \ne 0$ or $U \ne -2\epsilon_d$), we can imagine extending the applicability of the above formalism with the proviso that the potential scattering term is no longer forbidden.  With $V\ne 0$, the strong-coupling fixed point of p-hs is no longer stable{\cite{gonzalez}}.  For $0<r<r^*=0.375$, below a temperature scale{\cite{ingersent}} $T_{\chi'} \sim |\frac{V}{E^*}|^{\frac{1}{r}}T_{\chi}$ one flows{\cite{expl}} to an asymmetric strong coupling fixed point with entropy $S=0$ and phase shift{\cite{gonzalez}} $\delta = \pi sgn(-\epsilon)$ yielding $G = 0$.  
A summary of our results, including the peak width at finite $T$ is presented in Fig. \ref{figure3}; these are compared and contrasted with 1D results{\cite{kim}} where a one- 
or two-channel Kondo effect 
can occur depending on the range of interactions (and p-hs or the relevance of the $V$ term{\cite{1dstuff}}).          

To summarize, we have considered the idea of a quantum dot sandwiched between FL leads with a DOS (or equivalently hopping matrix elements) vanishing as a power law at the Fermi energy.  We recapture the low temperature dependence of the widths of the 1D Cbps ($r=-1+1/g$), while their height reaches $G=\frac{2e^2\cos^2(\frac{\pi r}{2})}{h}$ as $T\rightarrow 0$.  In contrast, Kondo physics drastically changes from the 1D case, as the phase shift varies continuously from $\delta = \frac{\pi}{2}$ to $\delta = \frac{\pi}{4}$ as the exponent $r$ changes from 0 to 0.5.  This is a signature of incomplete spin screening which coincides with a non-vanishing entropy{\cite{chen,gonzalez}} $S=2r$ln2 $(S(r=0)=0$: fully screened spin; $S(r=1/2)=$ln(2): free spin).  As in 1D{\cite{kim}}, p-h asymmetry matters leading to sharp peaks as a function of $V_G$. Observation of the disappearance of the Kondo signal  as a function of the dot-lead hopping at a critical Kondo coupling gives strong support for an underlying qcp.






We thank K. Ingersent for useful comments.  This work has been supported by CIAR, FQRNT and NSERC.


\end{document}